\documentclass[pra,twocolumn,superscriptaddress,showpacs,amsmath,amssymb]{revtex4}

\usepackage{bm}

\usepackage{graphicx}
\usepackage{latexsym}
\usepackage{amsmath}
\usepackage{amssymb}
\usepackage{amsfonts}
\usepackage{color}
\usepackage{bm}
\usepackage{verbatim}

\newcommand{\dd}[1]{\mathrm{d}#1\,}
\newcommand{\bra}[1]{\langle{#1}\rvert}
\newcommand{\ket}[1]{\lvert{#1}\rangle}

\DeclareMathOperator{\Tr}{Tr}

\begin{document}

\title{Lindblad equation approach for the full counting statistics of work and heat in driven quantum systems}

 \author{Mihail Silaev}
 \affiliation{ Low Temperature Laboratory, O.V. Lounasmaa Laboratory, Aalto
 University, P.O. Box 15100, FI-00076 AALTO, Finland }
 \affiliation{Institute for Physics of Microstructures
 RAS, 603950 Nizhny Novgorod, Russia}
 \author{Tero T. Heikkil\"a}
 \affiliation{ Low Temperature Laboratory, O.V. Lounasmaa Laboratory, Aalto
 University, P.O. Box 15100, FI-00076 AALTO, Finland }
 \affiliation{ Department of Physics, Nanoscience Center,
 University of Jyv\"askyl\"a, P.O. Box 35, FI-40014 University of
 Jyv\"askyl\"a, Finland }
 \author{Pauli Virtanen}
 \affiliation{ Low Temperature Laboratory, O.V. Lounasmaa Laboratory, Aalto
 University, P.O. Box 15100, FI-00076 AALTO, Finland }

\begin{abstract}
 We formulate the general approach based on the Lindblad equation  to
 calculate the full counting statistics of work and heat produced by driven quantum systems weakly coupled with a
 Markovian thermal bath. The approach can be applied to a
 wide class of dissipative quantum systems driven by an arbitrary
 force protocol.  We show the validity of general fluctuation relations and
  consider several generic examples. The
 possibilities of using calorimetric measurements to test the
 presence of coherence and entanglement in the open quantum systems are discussed.
  \end{abstract}

\pacs{} \maketitle

\section{Introduction}

The progress in experimental techniques provides a possibility to
study dissipative dynamics of mesoscopic and even quantum systems
where fluctuations play a significant
role\cite{ExperimentBio1,ExperimentPendulim,ExperimentPekola}.
Unlike for classical systems, the statistics of work is still not
well established in the quantum regime. Much attention has been
focused on derivations of the quantum versions of fluctuation
relations for open quantum systems
\cite{Campisi,Campisi1,Crooks,Esposito,CrooksQuantum,Mukamel,Roeck}.
In particular the fluctuation theorem was derived for an arbitrary
open quantum system in Ref.\cite{Campisi1}. Despite being a
powerful tool, such fluctuation theorems do not provide the
detailed information about the statistics of quantum fluctuations
which can be of great importance in driven quantum systems. Indeed
in quantum optics the sub-Poissonian statistics of photon counts
indicates the non-classical states of electromagnetic field
\cite{MandelWolfBook}. Observed first in the atomic resonance
fluorescence they have been seen later in many setups including
quantum dots, quantum wells and quantum point
contacts\cite{SubPoissonSources}.

Of particular interest is the recent experimental achievement to
realize an electrical circuit model of resonance fluorescence in a
single artificial atom\cite{Pashkin}. In this setup the atom was
represented by a superconducting qubit coupled to a transmission
line which can be considered as an effective thermal bath for the
open quantum system\cite{Carmichael}. In high contrast to the
optical devices the electrical circuit can have the temperature
compared to the qubit interlevel spacing\cite{Ustinov2007} so that
the work statistics can be strongly modified by thermal
fluctuations.

Recently the statistics of finite temperature work fluctuations in
a driven single qubit has been treated with the help of  a quantum
jump method \cite{Pekola}. In this paper we formulate an
alternative approach based on the generalized master equation
(GME) which can be applied to a wide class of open quantum systems
weakly coupled to the thermal bath. The applicability of the
suggested scheme relies on a quite general assumption that the
thermal bath is characterized by Markovian dynamics. In this case
the open quantum system can be described by the reduced density
matrix (DM) which satisfies the Lindblad equation which is a GME
with the Lindblad form of the dissipative operator
\cite{Lindblad}. We derive the generating functions which
determine the full counting statistics of the work and heat
exchange between the quantum system, thermal bath and the
classical source which implements an arbitrary drive protocol.
Previously the full counting statistics of charge and heat
transfer has been extensively investigated in non-equilibrium
mesoscopic systems\cite{Nazarov,Kinderman,Esposito}. The present
work is dedicated to develop an effective general approach to
calculate the fluctuations of work done by the external classical
source and the heat exchanged between the quantum system and
environment. We demonstrate the difference between the work and
heat statistics which becomes especially important for the small
exchanged amounts of energy. Several generic examples are
considered.

 The structure of the paper is as follows.
 In Sec. \ref{Sec:Formalism} we  develop
 the general formalism of Lindblad equation
 to calculate the full counting statistics of work and heat
 in driven quantum systems. In this section we
 demonstrate that the approximations made in order to trace out the environment variables keep the
 validity of  general fluctuation
 relations for the quantum work.
 In Sec.\ref{Sec:SingleQubit} the heat and work statistics in a single qubit
 at the finite temperature regime are considered. In Sec. \ref{Sec:LongTimeStat}
 we discuss an  analytical approach to calculate the long-time statistics of heat in the zero
 temperature limit. We apply this approach to several generic quantum systems such as a harmonic
 oscillator, a single qubit and two coupled qubits interacting with separate environments.
 Conclusions are given in Sec. \ref{Sec:Conclusion}.

 \section{Formalism and general fluctuation relations}
 \label{Sec:Formalism}
 \begin{figure}[!htb]
 \centerline{\includegraphics[width=1.0\linewidth]{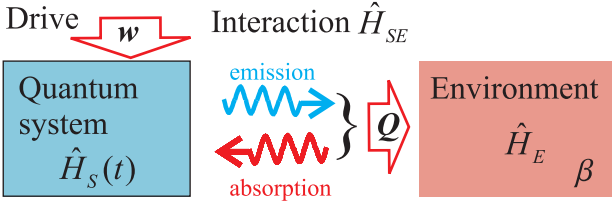}}
 \caption{\label{Fig:model} (Color online) Quantum system driven by an
external force performing the work $w$. The system interacts with
an environment (right) which is at equilibrium with inverse
temperature $\beta$. The interaction (middle) is described by an
emission and an absorption of energy from the environment. The
part of the work transferred to the environment and eventually
converted to heat is $Q$. }
 \end{figure}

We consider a system+environment model in which the system is
driven and it dissipates energy into an environment as shown
schematically in Fig. \ref{Fig:model}: $ \hat H(t) = \hat H_S(t) +
\hat H_{SE} + \hat H_E$, where $[\hat H_S(t), \hat H_E] = 0$. We
make a weak coupling approximation and neglect the contribution of
the interaction energy $\hat H_{SE}$ to the work done by the
system. In this case the work $w$ done by the external force is
defined according to the two-point projective measurement
\cite{Work,Kinderman} of the energy $\hat H_S+\hat H_E$.
 The full counting statistics of $w$ is determined by the generating function
\begin{equation}\label{Eq:GfDef1}
 G_{w}(u,t)= {\rm Tr} e^{iu\hat H_S(t)} \hat\rho (t,u),
\end{equation}
 which depends on the counting field $u$ and measurement time $t$.
 The DM $\hat\rho (t,u)$ is determined by the modified evolution operator
 \begin{eqnarray}\label{Eq:RoModif}
 \hat\rho (t,u)\equiv \hat U_{u/2} (t,0) \hat \rho (t=0,u) \hat U^\dagger_{-u/2}
 (t,0)\\ \nonumber
 \hat U_{u} (t,0) \equiv  e^{iu \hat H_E} \hat U (t,0) e^{-iu \hat
 H_E}
 \end{eqnarray}
 and the initial condition $\hat \rho (t=0,u)= e^{-iu\hat H_S(0)}\hat
 \rho_0$, where $\hat \rho_0$ is the DM of the
 system at the moment $t=0$ when we start to count the work.
 By writing (\ref{Eq:RoModif}) we assume
 that in the initial state the DM is diagonal in the basis of environment
 eigenstates
 $[\hat H_E,\hat \rho_0]=0$.

The work $w$ should be distinguished from the heat transferred to
the thermal bath, $Q$, that has a statistics given by the
generating function
 \cite{Campisi, Esposito}
 \begin{equation}\label{Eq:GfDef2}
 G_Q(u,t)= {\rm Tr}\; \hat\rho (t,u).
\end{equation}
 The DM dynamics is also given by Eqs. (\ref{Eq:RoModif}) but with a {\it different
 initial state}  $\hat \rho (t=0,u)= \hat \rho_0$.
  In general, the work and heat statistics can
  be drastically different.

 Let us take a simple form of the interaction
 term\cite{PuriBook} $\hat H_{SE}=\hat S^\dagger \hat R +c.c.$ and use the standard
 Born-Markov approximation assuming that the dynamics of the environment
 variable $\hat R$ is much faster than that of the system variable $\hat
 S$. In this case it is possible to trace out the $\hat R$
 variables to obtain the reduced DM of the quantum
 system $\hat{\tilde{\rho}}= {\rm Tr}_R \hat \rho$, which now determines the work generating
 functions
 (\ref{Eq:GfDef1},\ref{Eq:GfDef2}).
 Below we omit the tilde implying that DM is a
 reduced one.
  The reduced DM satisfies a Lindblad equation which depends on the system
variables only \cite{Esposito}
 \begin{equation}\label{Eq:Master0}
 \dot{\hat \rho}= \check A_u [\hat\rho],
\end{equation}
where $\check A_u [\hat\rho]= -i[\hat
H_S,\hat\rho]+\check{L}_u[\hat\rho]$ is a Liouvillian
superoperator and $\check{L}_u[\hat\rho]$ is a time-independent
Lindblad dissipative superoperator which describes the interaction
of the system with an environment\cite{remark1}
 \begin{eqnarray}\label{Eq:LindbladCFHO}
 \check{L}_u[\hat\rho]= \sum_j \Gamma_- e^{iu\nu_j} \hat S_j \hat\rho \hat S^\dagger_j + \Gamma_+ e^{-iu\nu_j}
 \hat S^\dagger_j \hat\rho \hat S_j -\\ \nonumber
 \Gamma_- \{\hat S^\dagger_j\hat S_j ,\hat\rho \} - \Gamma_+ \{\hat S_j\hat S^\dagger_j ,\hat\rho \}.
 \end{eqnarray}
 Here $\hat S_j (\hat S^\dagger_j)$ are lowering (rising) operators corresponding to the interlevel spacing $\nu_j$,
 $\{,\}$ is  an anti-commutator, and emission and absorption rates satisfy the detailed balance condition $\Gamma_+=e^{-\beta\nu_j}
 \Gamma_-$. By writing (\ref{Eq:LindbladCFHO}) we assume that the
 driving term is small enough to neglect the perturbation of level
 spacings $\nu_j$. In the opposite case the suggested approach can
 be modified to express the dissipative operator in the Floquet
 state representation\cite{Floquet}. Such approach allows
 describing
 Landau-Zehner processes in the dissipative systems\cite{PekolaLZ} but this issue
 is beyond the scope of the present paper.

 The general fluctuation relations can be obtained directly from Eqs.
 (\ref{Eq:GfDef1},\ref{Eq:Master0},\ref{Eq:LindbladCFHO}). Let us consider
 a detailed Crooks fluctuation relation\cite{Crooks}
  \begin{equation}\label{Eq:Crooks}
  P(w)/P^{tr}(-w) = e^{\beta w},
  \end{equation}
 which relates the probabilities of work $w$ and $-w$ done during forward and
  time-reversed driving protocols.
   The work distribution in time-reversed process is determined by the generating function
 introduced analogously to Eq. (\ref{Eq:GfDef1})
 \begin{equation}\label{Eq:GfDef1tr}
 G^{tr}_{w}(u,t)= {\rm Tr}\; e^{iu\hat H_S(0)} \hat\rho^{tr} (t,u).
 \end{equation}

  Indeed, the evolution of DM $\hat\rho^{tr} (t_1,u)$ when $t_1$ runs from $t$ to $0$ is determined
  by Eqs. (\ref{Eq:Master0},\ref{Eq:LindbladCFHO}) where $\check A^{tr}_u [\hat\rho]= i[\hat
H_S,\hat\rho]+\check{L}_u[\hat\rho]$. The initial condition is
$\hat \rho^{tr} (t_1=t,u)= e^{-iu\hat H_S(t)}\hat\rho_0 (t)$. The
Lindblad superoperator in Eq. (\ref{Eq:Master0}) is the same for
the forward and time-reversed protocols.
 To satisfy (\ref{Eq:Crooks}) it is enough to demonstrate  the
 generic symmetry relation
 \begin{equation}\label{Eq:Symmetry}
 G_w(u,t)=
 G^{tr}_w(i\beta-u,t)
 \end{equation}
 which holds provided that $\hat H_S(0) = \hat H_S(t)$ and the initial state for both the forward and time-reversed protocols is
 the equilibrium one $\hat\rho_0(t) = \hat\rho_0(0)= e^{-\beta \hat H_S(0)}/{\rm Tr} e^{-\beta \hat
 H_S(0)}$.
  The relation (\ref{Eq:Symmetry}) follows directly from
  Eq. (\ref{Eq:LindbladCFHO}) since due to the detailed balance and
permutation invariance of the trace the Lindblad superoperator
(\ref{Eq:LindbladCFHO}) satisfies
 \begin{equation}\label{}
  {\rm Tr}\; \hat\rho_1  \check{L}_{i\beta-u}[\hat\rho_2]  =
 {\rm Tr}\; \check{L}_{u}[\hat\rho_1] \hat\rho_2.
 \end{equation}
 Taking into account that $ {\rm Tr}\; \hat\rho_1  [H_S,\rho_2] = -{\rm Tr}\; [H_S,\rho_1] \hat\rho_2
 $ we get the relation for the Liouvillians of forward and
 time-reversed evolutions
 \begin{equation}\label{}
  {\rm Tr}\; \hat\rho_1  \check{A}_{u}[\hat\rho_2]  =
 {\rm Tr}\; \check{A}^{tr}_{i\beta-u}[\hat\rho_1] \hat\rho_2
 \end{equation}
 and consequently
 \begin{equation}\label{Eq:SymmetryA}
  {\rm Tr}\; \hat\rho_1  {\rm T} e^{\int_0^t A_u(t_1) dt_1 }[\hat\rho_2]  =
 {\rm Tr}\; {\rm \tilde{T}} e^{\int_0^t A^{tr}_u(t_1) dt_1 }[\hat\rho_1] \hat\rho_2
 \end{equation}
 where ${\rm T}$ and ${\rm \tilde{T}}$ are forward and
 time-reversed orderings.

 Using the formal solutions of Eq.(\ref{Eq:Master0}) the
 generating functions can be written as follows
 \begin{equation}\label{Eq:GfDef1Sol}
 G_{w}(u,t)= {\rm Tr} e^{iu\hat H_S(t)} {\rm T} e^{\int_0^t A_u(t_1) dt_1 } e^{-iu\hat H_S(0)}\hat
 \rho_0(0)
\end{equation}
 \begin{equation}\label{Eq:GfDef1trSol}
 G^{tr}_{w}(u,t)= {\rm Tr}\; e^{iu\hat H_S(0)} {\rm \tilde{T}} e^{\int_0^t A^{tr}_u(t_1) dt_1 } e^{-iu\hat H_S(t)}\hat\rho_0
 (t).
 \end{equation}

 Let us assume that both the forward and time-reversed
 evolution start from equilibrium and moreover $H_S(0)=H_S(t)=H_{S0}$ so that
 $\hat\rho_0 (\tau)  = e^{-\beta\hat H_{S0}}/ {\rm Tr e^{-\beta\hat H_{S0}}}$
 for $\tau=0,t$. Then the generic symmetry relation between
 generating functions (\ref{Eq:Symmetry})
 immediately follows from Eqs. (\ref{Eq:GfDef1Sol},
 \ref{Eq:GfDef1trSol})  and the relation (\ref{Eq:SymmetryA}).

 Since the heat $Q$ can be in practice measured either calorimetrically \cite{ExperimentPekola}
or by photon detection, statistics of both $w$ and $Q$ are in
general of interest. In certain limits the two statistics approach
each other, for example when the energy transfer to the
environment is large, or in certain low-temperature measurement
protocols. As $Q$ is not generically associated with fluctuation
relations such as Eqs. (\ref{Eq:Crooks}), they may fail to apply
even if the statistics otherwise approach each other (see Appendix
\ref{Appendix1} for discussion). In general it has been shown that
the average of the exponentiated heat depends on the details of
the driving protocol\cite{Talkner}. In Appendix \ref{Appendix2} we
calculate this quantity explicitly for a single qubit case and
demonstrate that it depends on the time of heat statistics
measurement.

 \section{Statistics of work and heat in a single qubit}
 \label{Sec:SingleQubit}

 Applying the general approach formulated above we
 study energy fluctuations in
  a driven two-level system described by the Hamiltonian
 \begin{equation}\label{Eq:Hamiltonian0}
 H_S(t)= \frac{\nu}{2} \hat\sigma_z+ \Omega\cos(\omega_d
 t)\hat\sigma_x,
 \end{equation}
where $\nu$ is a qubit level spacing and $\Omega$ is a pumping
intensity. The dissipation is described by the Lindblad operator
(\ref{Eq:LindbladCFHO}) with $\hat S=(\hat \sigma_x -
i\hat\sigma_y)/2$. To calculate the statistics of energy
fluctuations  we find the generating functions
(\ref{Eq:GfDef1},\ref{Eq:GfDef2}) by solving GME
(\ref{Eq:Master0}) numerically assuming that the system was in
thermal equilibrium at $t=0$.

\begin{figure}[!htb]
 \centerline{\includegraphics[width=1.0\linewidth]{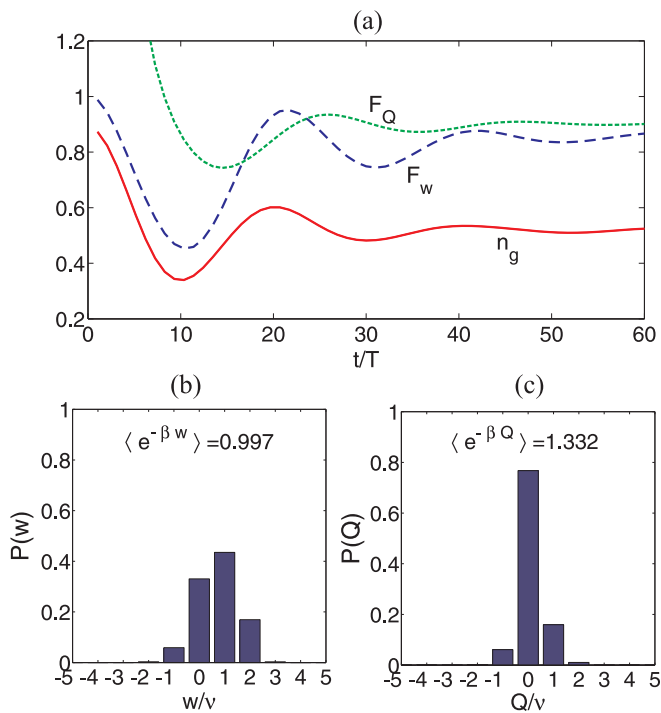}}
 \caption{\label{Fig:SingleQ} (Color online)
  (a) The ratio of the two lowest moments of work $F_w (t)$ and heat $F_Q(t)$ as functions
 of the measuring time normalized by the period $T=2\pi/\nu$. For comparison, the
 solid red line shows the Rabi oscillations of the ground state
 population. All dependencies are averaged over the period $T$. The temperature is $\beta=2/\nu$.
 (b,c) Distributions of work $w$ and heat $Q$ at the time moment
 $t=5T$. The calculated probabilities are $P_w(-2\nu:3\nu) = [ 0.0031, 0.0589, 0.3305, 0.4351, 0.1693,
 0.0030]$ and $P_Q(-2\nu:2\nu) =[ 0.0017, 0.0609, 0.7676, 0.1593, 0.0106]$.
 With numerical accuracy determined by the size of the counting field grid
 the work distribution satisfies Crooks and Jarzynski fluctuation
 relations. The panel (c) demonstrates lack of the fluctuation relations for the heat. }
 \end{figure}

 As noted in Ref. \cite{Pekola} the
ratio of the first two moments of work satisfies the linear
relation
 \begin{equation}\label{Eq:linear}
  \langle\Delta w^2\rangle /\langle w\rangle = \nu\coth (\nu/2T)
 \end{equation}
in the $\langle w\rangle\rightarrow 0$ limit realized for small
measuring times. This relation follows from (\ref{Eq:Crooks}) if
one assumes that only the probabilities $P(w=-\nu, 0,\nu)$ have
considerable values. At larger times the significant deviations
from Eq.(\ref{Eq:linear}) were found \cite{Pekola}. With our
approach we are able to consider the evolution of work statistics
at longer times in order to reveal the physical origin of such
behavior. We choose the Hamiltonian parameters
(\ref{Eq:Hamiltonian0}) as $\Omega=0.05\nu$, $\Gamma_-=0.007\nu$
and the temperature $\beta=2/\nu$. Starting from the equilibrium
state at $t=0$ we plot in Fig.\ref{Fig:SingleQ}a the time
dependance of the ratios $F_{w}= \tanh (\beta\nu/2)\langle\Delta
w^2\rangle /(\nu\langle w\rangle ) $ and $F_{Q}= \tanh
(\beta\nu/2)\langle\Delta Q^2\rangle /(\nu\langle Q\rangle )$. For
comparison we plot by solid red line the ground state population
of the qubit $n_g(t)$ which oscillates with a Rabi frequency
$\Omega$. The time dependence of
 $F_{w}(t)$ follows the Rabi oscillations of $n_g(t)$. On
the other hand the same dependence for the heat statistics
$F_{Q}(t)$ has phase-shifted oscillations with respect to
$n_g(t)$. Moreover in contrast to the work statistics we obtain
that the ratio $F_{Q}(t)$ diverges at $t\rightarrow 0$ due to the
fluctuating energy exchange between the system and the thermal
bath which exists even in the absence of the drive. In this case
the dispersion $\langle\Delta Q^2\rangle $ stays finite while
$\langle Q\rangle =0$.

More generally we obtain that the work distribution e.g. shown in
Fig.\ref{Fig:SingleQ}b obeys Jarzynski fluctuation
relation\cite{Jarzynski2007} $\langle e^{-\beta w}\rangle=1$
 and a particular form of Crooks fluctuation
relation $  P(w)/P(-w) = e^{\beta w}$  where $P(-w)=P^{tr}(-w)$
since the forward and inverse driving protocols for
Eq.(\ref{Fig:SingleQ}) coincide. On the other hand the heat
distribution shown in Fig.\ref{Fig:SingleQ}c does not obey the
fluctuation relations and hence the linear relation
(\ref{Eq:linear}). At elevated temperatures $\beta\ll 1/\nu$ the
Rabi oscillations of both the ground state population and the
moment ratios $F_{w,Q}(t)$ are strongly suppressed. In this limit
we obtain $F_w(t)=1$ at arbitrary times as well as $F_Q(t)=1$ at
$t \gg \Gamma^{-1}_-, \Omega^{-1}$.

 \section{Long-time statistics of heat in driven quantum systems}
  \label{Sec:LongTimeStat}

   Our approach based on the Lindblad equation allows analytical calculation of the long-term heat
    statistics in the quantum limit  $\beta\nu\gg 1$. In such
    regime we completely neglect the absorption events, assuming $\Gamma_+=0$ and $\Gamma_-=\Gamma$ in (\ref{Eq:LindbladCFHO}) so that
  \begin{equation}\label{Eq:LindbladCFHOT0}
 \check{L}_u[\hat\rho]= \Gamma e^{iu\nu} \hat S \hat\rho \hat S^\dagger  - \Gamma \{\hat S^\dagger\hat S ,\hat\rho \} .
 \end{equation} For the single interlevel spacing $\nu$ the heat transfer to the environment is
 proportional to the number of emitted photons in close analogy with quantum optics systems.
 In this sense the suggested approach can be employed to
 calculate the steady state statistics of photon counts without solving the non-stationary ME which in
 general is rather complicated even in the simplest case of a single qubit (two-level atom)\cite{MandelWolfBook}.

 In the steady state regime the statistics of work and heat coincide.
 In the long-time limit the amount of heat transferred to the environment is
 much larger than the internal energy of the quantum system. The
 latter thus can be neglected from Eq. (\ref{Eq:GfDef1}) and therefore
 we get that $G_w=G_Q$.

  In the steady state we calculate the moments of heat distribution measured over a large time interval $t\gg
\Gamma^{-1}, \Omega^{-1}$.
 In such a limit the generating function (\ref{Eq:GfDef2}) is determined by the
 Liouvillian eigenvalue $\lambda (u)$
 having the largest real part because an asymptotical solution of the
GME at $t\rightarrow\infty$ has the form $\rho (u,t)\sim
e^{t\lambda(u)}$. Hence the Taylor expansion
 \begin{equation}\label{Eq:Taylor1}
  \lambda (u)=-i\lambda_{1} u-\lambda_{2}u^2/2
 \end{equation}
 determines the leading order Gaussian distribution of $Q$ with the
 moments given by $\langle Q\rangle =   \lambda_{1} t$ and
 $\langle (\Delta Q)^2\rangle = \lambda_{2} t$.
  In order to find the  coefficients $\lambda_{1,2}$ in the general case we use an iterative
algorithm described in the Sec.\ref{SubSec:SingleQubit}. We start
however with calculating the steady state heat fluctuations in a
periodically driven harmonic oscillator.
 Recently the stochastic behavior
of damped harmonic oscillator was studied  with the help of the
quantum trajectories\cite{HarmonicOscillator}. Here we find
exactly the eigenvalue $\lambda(u)$ and hence the full counting
heat statistics.

\subsection{Harmonic oscillator }
 \label{SubSec:HarmonicOscillator}

   Let us consider the following Hamiltonian
 \begin{equation}\label{Eq:HamOsc}
  \hat H_S= \nu \hat a^\dagger\hat a +\Omega (\hat a^\dagger e^{-i\omega_dt}+\hat
  ae^{i\omega_dt}),
 \end{equation}
 where $\hat a^\dagger, \hat a$ are creation and annihilation operators
 and $\Omega$ is the drive force amplitude. The jump operators are
  $\hat S, \hat S^\dagger =\hat a, \hat a^\dagger$.
 In this case the ME
 has an exact steady state solution
 $\hat\rho_{ho}(t,u) = e^{\lambda(u) t}  \hat D^\dagger(\alpha)  |0\rangle\langle 0| \hat
 D(\alpha)$,
  where $\lambda(u)=\Gamma |\alpha|^2 (e^{iu\nu}-1)$ and
  $\hat D (\alpha)  = \exp ( \alpha \hat a^\dagger - \alpha^* \hat a)$ is a displacement transform
with $\alpha=2i\Omega e^{-i\omega_d t}/[\Gamma+2i(\nu-\omega_d)]$.
Using $\hat\rho_{ho}(t,u)$ to calculate the generating function
$G_Q(u,t)$ we obtain a Poissonian distribution $ P(n)= x^n
e^{-x}/n!$ of quantized heat $Q=n \nu$, where $x=\Gamma |\alpha|^2
t$.

  The obtained result is consistent with the well known one that
 a classical current source emits a coherent state of cavity modes
 which has a Poissonian distribution of photons \cite{Glauber}.
 In this model case each event of the heat transfer to the environment occurs at
 completely random times without any correlation with the previous
 ones. The linear response relation (\ref{Eq:linear})
 is satisfied exactly for arbitrary values of dissipation $\Gamma$ and pumping $\Omega$.

  \subsection{Single qubit}
  \label{SubSec:SingleQubit}

  A generic example which demonstrates significant deviations
 from linear response relation (\ref{Eq:linear}) is a periodically driven single qubit (two- level atom)
 where quantum correlations between subsequent photon emission
 events become important \cite{MandelWolfBook}.
     The Hamiltonian of such a system (\ref{Eq:Hamiltonian0})
  can be simplified in rotating wave approximation (RWA) to
 \cite{PuriBook} $\hat H_S=(\Omega\hat\sigma_x + \delta\hat\sigma_z)/2,$
where we introduce the detuning from resonance
$\delta=\nu-\omega_d$ and drive intensity $\Omega$.  The
dissipation is described by the Lindblad operator
(\ref{Eq:LindbladCFHOT0}) with
 $\hat S=(\hat\sigma_x- i\hat\sigma_y)/2$.

 The analytical expression of the full counting work statistics in a single qubit is not known.
 It is possible however to develop a general iterative approach to
 calculate moments of work distribution measured over a large time interval $t\gg \Gamma^{-1}, \Omega^{-1}$.
 As we discuss above in such a limit the
 statistics is determined by the Taylor expansion (\ref{Eq:Taylor1}) of the
  Liouvillian eigenvalue $\lambda (u)$
 having the largest real part.
 Note that the zeroth order term in $u$ is absent in (\ref{Eq:Taylor1}) due to the existence of
 a stationary solution at $u=0$ satisfying $\check A_0 [\hat\rho^{st}]
 =0$. In order to find Taylor coefficient $\lambda_{1,2}$ let us consider the expansion of the Liouvillian
 matrix $\check A_u$ up to the second order term
$ \check A_u [\hat\rho]=\check A_0[\hat\rho]+ \Gamma
[iu\nu-(u\nu)^2/2]
 \hat S\hat\rho \hat S^\dagger$
and search for the solution of the eigenvalue problem
 \begin{equation}\label{Eq:EigenValue}
 \check A_u [\hat\rho]=\lambda (u)\hat\rho
 \end{equation}
 in the form of an expansion
 $\hat\rho=\hat\rho^{st}+\hat\rho^{(1)}u +\hat\rho^{(2)}u^2$.
 At first  we retain the terms linear in $u$. We note that the
 transposed Liouvillian has one-dimensional kernel $\check A^T_0 [\hat\rho^{st}_T]=0$, where
$\hat\rho^{st}_T=\hat \sigma_0$ is just an identity matrix. This
property holds in general for trace preserving superoperators. We
take an inner product with $\hat\rho^{st}_T$ of both sides in
Eq.(\ref{Eq:EigenValue}) to obtain the solvability condition which
gives the value of $\lambda_{1}$. Substituting the obtained
$\lambda_{1}$ to Eq. (\ref{Eq:EigenValue}) again we find the
 correction to the DM $\hat\rho^{(1)}$.
 Next to determine $\lambda_{2}$ we include the terms $\sim u^2$ and repeat the
above procedure using the known values of the first order
corrections.

The heat transfer to the environment occurs via the emission of
monochromatic photons $Q=n\nu$, where $n$ is integer. Therefore
the heat statistics can be characterized by the Fano factor of
photon counts
 $F_{ph}= \sigma_Q^2/(\nu \langle Q\rangle
 )=\lambda_2/(\nu\lambda_1)$.
It determines the regimes of sub-Poissonian $F_{ph}<1$ and
super-Poissonian $F_{ph}>1$ quantum fluctuations.
 From the iteration scheme suggested above to calculate $\lambda_{1,2}$ we get the Fano factor for a single qubit
 $F_{ph}=1-2 \Omega^2(3\Gamma^2-4\delta^2)/(\Gamma^2+2\Omega^2+4\delta^2)^2$,
which corresponds to the sub-Poissonian regime for small detuning
$\delta<\sqrt{3}\Gamma/2$ and otherwise to the super-Poissonian
one. This result agrees with the statistics of photon counts in
the resonance fluorescence of a two-level atom where the
non-classical sub-Poissonian statistics originates from photon
antibunching \cite{Mandel,MandelWolfBook,Cook}.

 \subsection{Super-Poissonian heat fluctuations and entanglement of two coupled qubits}
 \label{SubSec:CoupledQubits}

 As discussed above the dependencies shown in Fig.(\ref{Fig:SingleQ}) clearly
demonstrate that the peculiarities of heat and work statistics can
be considered as a signature of the finite coherence in the open
quantum system. To obtain more evidence of that we proceed to
investigate whether such a purely quantum property as the
entanglement between two coupled qubits can be tested by measuring
the heat produced by the quantum system.

It is well known that the entanglement between qubits can be
induced by dissipative
processes\cite{EntanglementEnvironment,Sorin}.
 The photon statistics in this system with incoherent pumping was
considered in Ref. \onlinecite{delValle}. The conclusion in that
case is that the photon statistics is always sub-Poissonian
regardless of entanglement. Here we demonstrate that for the case
of coherent pumping the statistics of heat transferred to the
environment has distinctive features in a subset of the possible
entangled states. In general, however, we note that there is no
exact correspondence between the statistics of emitted noise and
the presence of steady state entanglement in the quantum system.
Even a monochromatically driven single qubit can emit both super-
and sub-Poissonian distribution of photons \cite{Cook}. Indeed,
two-time correlation functions, such as those for the work or heat
statistics, are not solely a property of the steady state density
matrix, but are determined also by the transient dynamics of the
quantum system between subsequent emission events. However, when
consideration is restricted to Hamiltonians of a particular form,
such as the setup discussed below, correspondence can arise.

We consider the Hamiltonian of two coupled identical two-level
systems which can model the system of inductively coupled phase
qubits\cite{CoupledQubits}. For simplicity we focus on the quantum
fluctuations of $Q$ in the zero temperature limit neglecting  the
absorption events in (\ref{Eq:LindbladCFHO}) completely. Hence
this system can be considered as a model
 of coupled atoms. Its RWA Hamiltonian in case of resonant driving
 has the form
 \begin{equation}\label{Eq:HamiltonianRWA}
 \hat H_S=\frac{1}{2}\sum_{j=1,2}\Omega  \hat\sigma^{(j)}_x+
 \omega_{xx}\left(\hat\sigma^{(1)}_x\hat\sigma^{(2)}_x+\hat\sigma^{(1)}_y\hat\sigma^{(2)}_y\right).
\end{equation}
  where $\omega_{xx}$ is the qubit coupling and $\Omega$ is a
pumping intensity. The parameters of the qubits are assumed to be
identical. The dissipation is described by the ME
(\ref{Eq:Master0}) with a Lindblad operator
(\ref{Eq:LindbladCFHO}) with $\hat S_j=(\hat \sigma^{(j)}_x -
i\hat\sigma^{(j)}_y)/2$. This form of the dissipation operator
corresponds to the qubits interacting with separate thermal baths.
 The steady state entanglement in this system generated by the interaction with dissipative environment
 was discussed in Ref. \cite{Sorin}.

 To describe the statistics of work done by each qubit in separate we
 introduce the generating function which depends on
 two quantum fields $G_Q=G_Q(u_1,u_2,t)$.
  By setting $u_1=u_2$ we obtain the
statistics in the double channel which does not distinguish
between the heat produced by each of the qubits. The single
channel generating function e.g. for the first qubit is given by
$G_{Q1}(u_1,t) =G_Q(u_1,0,t)$.

\begin{figure}[!htb]
 \centerline{\includegraphics[width=1.0\linewidth]{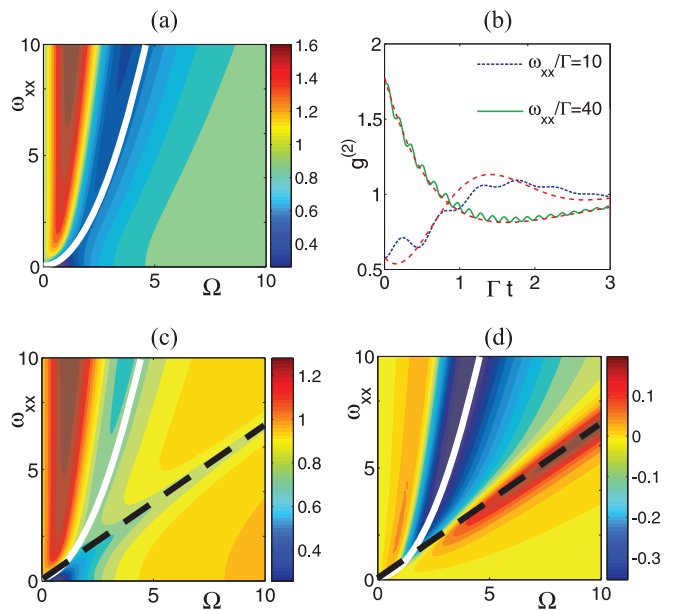}}
 \caption{\label{Fig:Statistics} (Color online)
  Statistics of heat produced by two coupled qubits.
 (a) Fano factor in a double channel $F_{ph}=\sigma_Q^2/(\nu \langle Q\rangle )$ as a function of interaction
  strength $\omega_{xx}$ and pumping $\Omega$.
 (b) Second order correlation functions - the probabilities of sequential photon emission by the two-qubit system
    compared to the effective three level approximation model
    (dashed curves). The parameters are $\Omega=5$, $\Gamma=1$ and
    $\omega_{xx}=10$ (solid curve), $\omega_{xx}=40$ (dotted curve).
  (c) Correlation function
  $C_{12}/(\nu^2\Gamma t)$ of heat produced by each of the qubits in separate.
  (d) Fano factor of heat released by one of the qubits
  $F_{ph1}=2\sigma_{Q1}^2/(\nu \langle Q\rangle )$.
   The white solid curves show the boundary
     between the steady state entangled and non-entangled regimes. The relaxation rate is $\Gamma=1$.
   The super-Poissonian regimes both in
    the double and the single channels correspond to finite entanglement. Dashed black line (b,c) shows the resonance condition
    $\Omega=\sqrt{2}\omega_{xx}$.
    }
 \end{figure}

  For further calculations we use the stationary solution of
 GME given by the kernel of the Liouvillian superoperator $\check A_0 [\hat\rho^{st}]=0$
  obtained in Ref. \cite{Sorin}.
To find the moments of work we calculate the coefficients in the
expansion of the eigenvalue (\ref{Eq:EigenValue})
 \begin{equation}\label{Eq:Expansion2}
  \lambda(u_1,u_2) = -i\lambda_1 (u_1+u_2)- \frac{\lambda_2}{2}( u_1^2+ u_2^2)-\lambda_{12} u_1u_2.
  \end{equation}
 By using the same iterative
algorithm as for the single qubit we find
 an analytical expression for the Fano factor $F_{ph}= \sigma^2_Q/(\nu\langle Q\rangle)$
 which is too long to be written here.
 In Fig.\ref{Fig:Statistics}a we plot $F_{ph}$
 as a function of parameters $\omega_{xx}, \Omega$ at fixed $\Gamma=1$.
  There is a region on this plane, where
  $F_{ph}>1$ indicating super-Poissonian statistics.
 The possibility of a super-Poissonian regime is
 explained by the positive correlation between
 works extracted from two qubits (see below).
 It is realized for a strong coupling between
  the qubits and small dissipation $\omega_{xx}\gg\Omega\gg
  \Gamma$ and describes bunching of emitted photons which is in contrast to the
 antibunching in a single qubit resonant fluorescence.

  On the qualitative level the photon bunching and super-Poissonian statistics
  result from the steady state entanglement between the coupled qubits\cite{Sorin}.
  Let us transform the steady state DM to the energy basis of the
  effective Hamiltonian (\ref{Eq:HamiltonianRWA}). For
  $\omega_{xx}\gg \Omega$ the energies are given by $\varepsilon_1=0$,
  $\varepsilon_{2,3}=[\omega_{xx}\mp\sqrt{4\Omega^2+\omega_{xx}^2}]/2$,
  $\varepsilon_4=-\omega_{xx}$.
 In the regime $\omega_{xx}\gg\Omega$ the basis is formed by the Bell
 states. In particular
 $\Phi^{\pm}= |0\rangle_1 |0\rangle_2 \pm |1\rangle_1 |1\rangle_2$
 correspond to the first two levels separated by the smallest Rabi frequency
 $\Omega_R=\varepsilon_1-\varepsilon_2\approx \Omega^2/\omega_{xx}$.
  In the DM we neglect the small terms proportional to $\Omega/\omega_{xx}$
 and get the following non-zero elements:
 \begin{eqnarray} \label{Eq:3LevelSteadySt}
 \rho_{11}=\rho_{22} = 1/2-\Omega_R^2/4X, \;
 \\ \nonumber
 \rho_{12}=\rho^*_{21}=[\Gamma^2-i\Gamma\Omega_R]/2X,
 \end{eqnarray}
 where $X=\Omega_R^2+\Gamma^2$ and $ \rho_{33}=\rho_{44} = 1/2-\rho_{11}$. Hence at $\Gamma\gg\Omega_R$
 the most populated are the Bell states $\Phi^{\pm}$.

 In this case the emission of a photon by one of the qubits
   can be considered as a projective measurement \cite{Pekola} which drives the
  other qubit to the excited state and hence triggers
  an  emission of the next photon once this qubit relaxes.
  In Fig.\ref{Fig:Statistics}a we plot by a white solid line the
  boundary $\omega_{xx}= \Omega^2/2\Gamma$ which separates
  the region with finite steady state entanglement $\omega_{xx}> \Omega^2/2\Gamma$
  from that of non-entangles states\cite{Sorin}. The super-Poisson
  statistics of work is realized only for the entangled states and
  can be considered as a possible experimental signature of entanglement.

   To obtain a quantitative description of the photon bunching
 we consider the second order correlation function of photon
emission intensity\cite{MandelWolfBook}
 $ g^{(2)}(t) $ plotted in Fig.\ref{Fig:Statistics}d for the
parameters $\Omega/\Gamma=5$ and $\omega_{xx}/\Gamma=10$ (blue
dotted curve) and $\omega_{xx}/\Gamma=40$ (green solid curve). It
demonstrates the
 photon bunching $g^{(2)}(0)\geq g^{(2)}(t)$ at large values of
 $\omega_{xx}\gg \Omega\gg \Gamma$. The correlator has rapid oscillations with a frequency
  determined by the coupling $\omega_{xx}$ superimposed over the
  smooth behavior.

  In order to get more insight in the details of the curves shown
  in Fig.(\ref{Fig:Statistics})b we obtain an analytical description of the
  correlations averaged over the fast oscillations neglecting the small elements of the DM
  proportional to $\Omega/\omega_{xx}$ and $\Gamma/\omega_{xx}$.
  This approximation results in
  an effective three-level ansatz for the DM with non-zero elements
  $\tilde{\rho}_{ij}= \rho_{ij}$ for $i,j=1,2$ and $\tilde{\rho}_{33} =
  \rho_{33}+\rho_{44}$. The steady state is determined by (\ref{Eq:3LevelSteadySt}).
  The resulting ME has a Hamiltonian with
  non-zero matrix elements $H_{11}=-\Omega_R/2$ and
  $H_{22}=\Omega_R/2$. The elements $H_{33},H_{44}$ drop from the ME
  for the above ansatz of DM.
  The Lindblad superoperator
  (\ref{Eq:LindbladCFHOT0}) has $\hat S=\hat S_3$ which is a $3\times 3$ matrix
  with non-zero elements
  $(\hat S_3)_{13}=(\hat S_3)_{23}=i\sqrt{\Gamma/2}$ and $(\hat S_3)_{31}=-(\hat S_3)_{32}=i\sqrt{\Gamma}$.
  The second order correlator
  $ \langle g^{(2)}\rangle(t)= \langle \hat S_3^\dagger  \hat S_3^\dagger (t) \hat S_3 (t) \hat S_3 \rangle/
  \langle \hat S_3^\dagger  \hat S_3 \rangle^2$
  gives the correlator $g^{(2)}(t)$ of the full system
  averaged over the period of fast oscillations $\omega^{-1}_{xx}$ . Most
  importantly the operator $\hat S_3$ allows for
  the two simultaneous quantum jumps from the steady state
  by (\ref{Eq:3LevelSteadySt}) since $\hat S^2_3 \hat\rho^{st} \hat S^{^\dagger 2}_3\neq 0$.
  In this model the two photons can be emitted
  simultaneously which gives the finite value of $g^{(2)}(t=0)$.

  It is possible to
  find $\langle g^{(2)}\rangle(t)$ analytically by solving the ME with $u=0$
  and the initial state after the
  first quantum jump given by $\hat S_3 \hat\rho^{st} \hat S^\dagger_3$.
  Indeed from the three-level ME we see that the equations for the
  components $\tilde{\rho}_{11}$ and $\tilde{\rho}_{22}$ coincide and thus we can reduce
  the problem to the second order ME. It can be solved with arbitrary initial
  conditions yielding the correlator
  \begin{equation}\label{Eq:AverageG2}
  \langle g^{(2)}\rangle (t)=  1+ e^{-\Gamma t}\left[\frac{(x^2-1)}{2}\cos(\Omega_R t)-x\sin(\Omega_R t)\right],
  \end{equation}
  where we introduce $x=\Gamma/\Omega_R$.
  The function $\langle g^{(2)}\rangle (t)$ is plotted in Fig. (\ref{Fig:Statistics})b by dashed lines.
  It has a damped oscillating behavior determined by the
  competition of an effective Rabi frequency $\Omega_R$ and the damping $\Gamma$.
  In high contrast to the single qubit case the two-point
  correlation function can have a maximum at $t=0$ at large values of
  $\omega_{xx}$ which corresponds to the bunching of photons produced by the
  entangled states of the two qubit system.


 Next we use the expansion (\ref{Eq:Expansion2}) to find the correlation function of heat produced by
each of the qubits
 \begin{equation}\label{Eq:Correlation}
 C_{12}= \langle Q_1Q_2\rangle-\langle Q_1\rangle\langle Q_2\rangle= \lambda_{12}
 t.
 \end{equation}
 In Fig. \ref{Fig:Statistics}(c) one
can see two peaks of $C_{12}$ as a function of parameters
$\Omega,\omega_{xx}$. One of them corresponds to non-zero
entanglement between the qubits and produces super-Poisson
statistics of work. The second peak is located at
$\omega_{xx}=\Omega/\sqrt{2}$ as indicated by the dotted black
line in Fig. \ref{Fig:Statistics}(c).
 For such parameters the spectrum of effective RWA Hamiltonian
 (\ref{Eq:HamiltonianRWA}) is degenerate since the two levels
 coincide, $\varepsilon_2=\varepsilon_4$. The subspace of
 eigenstates for this level is given by
 $\psi_{24}=c_2\psi_2+c_4\psi_4$,
 where $c_{2,4}$ are arbitrary coefficients, and the orthogonal
 basis functions are $\psi_2=(\sqrt{2},1,1,\sqrt{2})/\sqrt{6}$ and
 $\psi_4=(0,-1,1,0)/\sqrt{2}$.
 To understand the nature of positive correlation $C_{12}$,
 let us consider the case of small relaxation $\Omega,\omega_{xx}\gg
 \Gamma$ when the DM is
 diagonal $\hat\rho=\left( |11\rangle\langle 11| + |00\rangle\langle 00|+ |01\rangle\langle 01|+
  |00\rangle\langle 00|\right)/4$ . After the photon emission for example, by the  first qubit the
 DM is projected to $\hat\rho = \hat S_1 \hat\rho_0 \hat S^\dagger_1= (|01\rangle\langle 01|+
  |00\rangle\langle 00|)/4$ because the first qubit is bound to be in the ground state.
  Taking the time average of the DM we find that during the subsequent evolution the first term
  produces a contribution $|01\rangle\langle 01|\rightarrow |\psi_{24}\rangle\langle \psi_{24}|$
  with  $c_2=1/\sqrt{6}$ and $c_4=1/\sqrt{2}$ so that
  in this case the amplitude of $\psi_{24}$ is asymmetric.  At the
  same time the second term $|00\rangle\langle 00|$ evolves into
  a symmetric time-averaged population of the qubits. That is  the probability for the second qubit to
  be in the excited state and hence to emit a photon is larger
  than for the first one. Thus the peak of $C_{12}$ is
  accompanied by the drop of the single channel Fano factor as
  shown in Fig. \ref{Fig:Statistics}(d) so that the statistics of
  work in double channel is sub-Poissonian,
 $F_{ph}<1$ [Fig. \ref{Fig:Statistics}(a)].


   \section{Conclusion}
   \label{Sec:Conclusion}

To conclude, we have developed a formalism of Lindblad equation to
calculate the statistics of energy fluctuations in dissipative
quantum systems driven by an external force. We introduced
generating functions of work and heat exchanged between the
system, the classical driving source and the thermal bath which
can be found by solving the Lindblad equation. The general
fluctuation relations are shown to be valid for the resulting work
statistics. Applying this formalism to the generic examples of a
harmonic oscillator, a single and two coupled qubits we have shown
that calorimetric measurements of heat statistics can indicate the
presence of finite coherence and entanglement in the open quantum
systems.

It is our pleasure to thank Prof. Jukka Pekola for discussions.
This work was supported by the Academy of Finland, the European
Research Council (Grant No. 240362-Heattronics), and the EU-FP 7
INFERNOS (Grant No. 308850) programs.

\appendix

\section{Measuring work statistics at zero temperature}
\label{Appendix1}

At zero temperature, it is possible to determine the long-time
statistics $G_w(u,t\to\infty)$ of the total work $w$ by a
measurement of the statistics $G_Q(u,t\to\infty)$ of the heat
dissipated to the environment $Q$. The measurement scheme is to
drive the system only during the time interval $0<t<t_1$, after
which the system Hamiltonian is made constant and to coincide with
the initial state, $H_S(t) = H_S(0)$ for $t>t_1$. In this case,
the internal energy stored in the system is eventually emitted to
the zero-temperature environment, \cite{Pekola} and one finds
$G_w(u,t\to\infty) = G_Q(u,t\to\infty)$.

This correspondence can be proven directly using our master
equation formulation. We start with the observation that for a
given constant $u$, we have at time $t>t_1$,
\begin{align}
  \lim_{\beta\to\infty}\check{A}_u(t)\{P_0\} = 0
  \,,
\end{align}
for the ground state density matrix $P_0=\ket{0}\bra{0}$. Indeed,
$\hat{S}P_0=P_0\hat{S}^\dagger=0$. Therefore, $P_0$ is a steady
state of the $u$-dependent time evolution for $\beta\to\infty$,
and we find for $\beta,t\to\infty$,
\begin{align}
  {\cal T}e^{\int_0^t\dd{t'}\check{A}_u(t')}\{P_0\}
  \to
  f(u,t) P_0
  \,,
\end{align}
where $f(u,t)$ is a scalar function. The initial state for $G_w$
is $\rho_0 = e^{-(iu+\beta)H_S(0)}/\Tr[e^{-\beta H_S(0)}] \to
e^{-iu\epsilon_0}P_0$ for $\beta\to\infty$, where $\varepsilon_0$
is a ground state energy. It then immediately follows that for
$\beta,t\to\infty$,
\begin{align}
  G_w(u,t)
  \to
  e^{-iu\epsilon_0}
  \Tr[
  e^{i u H_S(0)} f(u,t) P_0
  ]
  =
  f(u,t)
  \,.
\end{align}
On the other hand, we have for the dissipated work, for
$\beta,t\to\infty$,
\begin{align}
  G_Q(u,t)
  =
  \Tr {\cal T}e^{\int_0^t\dd{t'}\check{A}_u(t')}\{P_0\}
  \to
  f(u,t)
  \,.
\end{align}
Therefore, in this measurement scheme we find
 $G_w(u,t\to\infty)=G_Q(u,t\to\infty)$, which is also the result
one would expect based on physical arguments.

Note that above $u$ is taken as a $\beta$-independent constant
while taking the limit $\beta\to\infty$. Therefore, although we
find that $G$ converges pointwise to $G_w$, this does \emph{not}
imply that the dissipated work satisfies the Jarzynski equality.
Indeed, while $G_w(i\beta,t)=1$, the result above does not imply
that $\lim_{t\to\infty, \beta\to\infty}G_Q(i\beta,t)=1$.

 \section{Lack of fluctuation relations for $Q$}
\label{Appendix2}

The lack of {\it universal} fluctuation relations for the heat $Q$
can be easily demonstrated based on the master equation approach.
This can be seen in the model of an non-driven qubit coupled to an
environment. The corresponding Liouvillian is
\begin{align}
  \check{A}_u(t)\{\rho\}
  &=
  -i[ \frac{\nu}{2}\sigma_z, \rho]
  +
  \Gamma \Bigl( e^{iu\nu} \hat{S} \rho \hat{S}^\dagger - \frac{1}{2}\{\hat{S}^\dagger\hat{S},\rho\}\Bigr)
  \\\notag
  &+
  \Gamma e^{-\beta\nu} \Bigl( e^{-iu\nu} \hat{S}^\dagger \rho \hat{S} - \frac{1}{2}\{\hat{S}\hat{S}^\dagger,\rho\}\Bigr)
  \,.
\end{align}
Solving the time dependence of the $u$-dependent master equation,
we find the generating function for the dissipated work
\begin{align}
  \label{eq:example-G-func}
  G_Q(u,t)
  &=
  \frac{
    \cos(u\nu) + \cosh(\beta\nu)
  }{
    1 + \cosh(\beta\nu)
  }
  \\\notag
  &
  -
  e^{-(e^{-\beta\nu}+1) \Gamma t}
  \frac{e^{(\beta-i u)\nu} (e^{i u \nu} - 1)^2}{(e^{\beta\nu}+1)^2}
  \,.
\end{align}
This implies that the heat only obtains values $Q=\pm\nu,\,0$, as
can be expected based on the limited internal energy of the qubit.

The above implies that the expectation value $\langle{e^{-\beta
    Q(t)}}\rangle=G_Q(i\beta,t)$ is time dependent.  Because this occurs
at equilibrium with all Hamiltonians time independent, there can
be no fluctuation relation of the form $\langle{e^{-\beta
    Q(t)}}\rangle=e^{-\beta\Delta F}$ where $\Delta F$ is a difference
between two equilibrium free energies. This result is in agreement
with the general non-equilibrium equality for the heat derived in
Ref. \onlinecite{Talkner}.

Moreover, Eq.~\eqref{eq:example-G-func} serves as a simple example
for the above discussion of the zero-temperature limit. Indeed,
for given fixed $u$ we have
 $\lim_{\beta,t\to\infty}G_Q(u,t)=1=G_w(u,t)$. However, at the same time,
$\lim_{\beta,t\to\infty}G_Q(i\beta,t)=2$.


\begin{thebibliography}{99}
\bibitem{remark1}
 The counting factors $e^{\pm iu\nu}$ in Eq.(\ref{Eq:LindbladCFHO}) arise from
 the generalized correlation functions of the environment
 variables $\int d\omega \delta (\omega-\nu) e^{iu\omega} {\rm Tr_R}[ \hat\rho_0 \hat R(\omega) \hat
 R(-\omega)]$. This approach can be applied to systems with
 multiple eigenfrequencies $\nu_i$. It is valid in case of the weak drive
 which means that the drive does not change the eigenfrequencies. 
\end{thebibliography}
\end{document}